\documentstyle[twocolumn,epsfig,aps]{revtex}
\begin{document}
\draft
\title{Quantum entanglement using trapped atomic spins }
\author{L. You and M. S. Chapman}
\address{School of Physics, Georgia Institute of Technology, \\
Atlanta, GA 30332-0430}
\date{\today}
\maketitle

\begin{abstract}
We propose an implementation for quantum logic and computing using trapped
atomic spins of two different species, interacting via direct magnetic
spin-spin interaction. In this scheme, the spins (electronic or nuclear) of
distantly spaced trapped neutral atoms serve as the qubit arrays for quantum
information processing and storage, and the controlled interaction between
two spins, as required for universal quantum computing, is
implemented in a three step process that involves state swapping with a
movable auxiliary spin.
\end{abstract}

\pacs{03.67.Lx, 32.80.Pj, 34.90.+q, 67.57.Lm}

\narrowtext

The field of quantum computing has advanced remarkably in the few years
since Shor \cite{shor} presented his quantum algorithm for efficient prime
factorization of very large numbers, potentially providing an exponential
speed up over the fastest known classical algorithm. Because much of today's
crytpography \cite{rsa} relies on the presumed difficulty of factoring large numbers,
Shor's discovery has important implications to data encryption technology
and has stimulated much work in the field of quantum information.

Motivated by this and other theoretical developments, there is much interest
in identifying and realizing experimental systems capable of generating
large-scale quantum entanglement. In atomic systems, there have been
several recent proposals using trapped ions/atoms and cavity QED systems
\cite{Cirac,Cirac2,km,atac,caves}.
Indeed, atomic systems capable of entangling two qubits have already been
realized in some of these systems \cite{nist,kimble}.
A common element in most of these
proposals is that the qubits are stored in distinguishably trapped
atoms/ions. The proposals differ principally in the nature of the atom-atom
interaction (either phonons, photons, collisional, and induced
electric-dipole moments) and in the way that these interactions are
controlled.

In this paper, we propose an implementation of a quantum logic scheme
utilizing the direct magnetic spin-spin interaction between individually
trapped neutral atoms. The qubits of this system are stored in the
long-lived hyperfine ground states of atoms, and coherent control of the
spin-spin interactions is accomplished by controlling inter-atomic spacings.
Our proposal is distinctive in that (1) the magnetic spin-spin interaction
used to create the inter-atom entanglement is virtually decoherence-free
and (2) atom-atom interactions are mediated via a movable `header' atom
which serves to transfer quantum information from one qubit to another (see
Fig. 1).  The header atom can in fact be a different species, and hence, in
contrast to \cite{km,caves,zoller}, the atom trapping potentials are not
required to be spin-dependent in order to maintain trap distinguishability
for small atom separations and can be realized with far-detuned laser beams.
This latter distinction is important because near-resonant laser traps are a
significant source of decoherence.

We begin by considering in detail the inter-atomic potential between two
neutral atoms separated by an inter-nuclear distance $\vec{R}$. For
the moment, we assume two spin 1/2 alkali atoms and momentarily neglect the
hyperfine interactions. The potential can be written as the sum of three
terms \cite{stoof}.
\[
V(\vec{R})=V_{T}(\vec{R}){\cal P}_{T}+V_{S}(\vec{R}){\cal P}_{S}+V_{D},
\]
$V_{T}$ and $V_{S}$ are the (electronic) spin triplet and singlet
potentials, respectively. $V_{D}$ represents the long range direct magnetic
dipole interaction between two atoms. ${\cal P}_{T}$ and ${\cal P}_{S}$ are
the projection operators into the total electronic subspace 1 (triplet) and
0 (singlet). The difference between $V_{T}$ and $V_{S}$ represents the
exchange interaction, which is typically most important when $R$ is less than the
LeRoy radius $R_{0}$ ($\lesssim $ $40a_{0}$ for two identical alkali atoms).
For two different atom species, the exchange interaction is considerably
suppressed beyond the contact limit (a few $a_0$).
In the long range limit both $V_{T}$ and $V_{S}$ are dominated by the
van de Waals term $-C_{6}/R^{6}$ \cite{mircea}.

At low energies, we can re-express the first two terms of the potential by
writing the spin triplet and singlet potentials in terms of the scattering
lengths, $a_{T}$ and $a_{S}$
\[
V_{\nu }(\vec{R})={\frac{4\pi \hbar ^{2}}{M}}a_{\nu }\delta (\vec{R}),%
\hspace{0.5in}\nu =T,S.
\]
and explicitly evaluating the projection operators to yield
\begin{eqnarray}
V(\vec{R})=&&{\frac{4\pi \hbar ^{2}}{M}}({\frac{3}{4}}a_{T}+{\frac{1}{4}}%
a_{S})I\,\delta (\vec{R}) \nonumber \\
+&&{\frac{4\pi \hbar ^{2}}{M}}(a_{T}-a_{S}){\frac{1}{4}}\vec{\sigma}%
_{1}\cdot \vec{\sigma}_{2}\,\delta (\vec{R})+V_{D}.
\label{op}
\end{eqnarray}
where $\vec{\sigma}_{v}$
is electron Pauli spin operators \cite{Julio}.

At this juncture, we point out that the recent Innsbruck proposal \cite
{zoller} employs the close-range part of the potential represented in the
first line of Eq. (\ref{op}) in a type of ``controlled collision". In our scheme
we will use the spin-dependent interaction, i.e. the long range atomic
magnetic interaction represented in the last term of Eq. (\ref{op}). The
proposal of Brennen, {\it et. al.} \cite{caves} relies on the near-resonant
electric dipole interaction (not present here in the ground state
Hamiltonian).

It is convenient to re-express the second term of Eq. (\ref{op}) by
assuming that the two interacting atoms (denoted by subscripts $q$ and $h$)
are harmonically bound in cylindrically symmetric traps with characteristic
radial and axial sizes: $a_{qr}$, $a_{hr}$, $a_{qz}$, and $a_{hr}$ and
furthermore that the atoms occupy the ground states of their respective
traps $|0\rangle =|0\rangle_{q}|0\rangle_{h}$. In this case, we obtain
\begin{eqnarray}
{\cal J}_{E} &=&\langle 0|{\frac{4\pi \hbar ^{2}}{M}}(a_{T}-a_{S})\delta (%
\vec{r}_{q}-\vec{r}_{h}-z_{0}\hat{z})|0\rangle   \nonumber \\
&=&{\frac{4}{\sqrt{2\pi }}}(a_{T}-a_{S}){a^2\over a_r^2}
{\hbar \omega \over a_z}\,e^{-{\frac{z_{0}^{2}}{%
2a_{z}^{2}}}},  \nonumber
\end{eqnarray}
for a reference harmonic trap frequency $\hbar \omega $ with
ground state size $a$. We have used $\vec{r}_{q}$ and $\vec{r}_{h}$ for the nuclear
coordinates of the atoms with respect to their own trap centers, which are
displaced by $\vec{d}=(0,0,z_{0})$, and we have defined
$
a_{\nu }=\sqrt{a_{q\nu }^{2}+a_{h\nu }^{2}}, (\nu =r,z).
$
It's important to recognize that ${\cal J}_{E}$ decays exponentially with
the nominal atom-atom separation, $z_{0}$.

The last term in Eq. (\ref{op}),
$V_{D}$, contains three separate terms corresponding to electron-electron,
electron-nuclear, and nuclear-nuclear magnetic dipole interactions.
Between alkali atoms, the strongest is the electron dipole interaction
\[
V_{D}^{ee}={\frac{\mu _{e}^{2}}{R^{3}}}[\vec{\sigma}_{q}\cdot \vec{\sigma}%
_{h}-3(\hat{R}\cdot \vec{\sigma}_{q})(\vec{\sigma}_{h}\cdot \hat{R})],
\]
where $\mu _{e}$ is the Bohr magneton. The strength of this
interaction is
\[
\gamma _{e}(R)={\frac{\mu _{e}^{2}}{R^{3}}}\approx 5\times 10^{11}\left( {%
\frac{a_{0}}{R}}\right) ^{3}({\rm Hz}),
\]
while $\gamma_{en}(R)$ (electron-nuclear) and
$\gamma_{n}(R)$ (nuclear-nuclear) are about $10^{-3}$ and $10^{-6}$
times smaller respectively. Therefore one may effectively write the
spin-dependent interaction Hamiltonian as
\[
{\cal H}={\cal J}_{E}(z_{0})\,\vec{\sigma}_{q}\cdot \vec{\sigma}_{h}+\gamma
_{e}(R)[\vec{\sigma}_{q}\cdot \vec{\sigma}_{h}-3(\hat{R}\cdot \vec{\sigma}%
_{q})(\vec{\sigma}_{h}\cdot \hat{R})].
\]
Typically, we will have $\gamma _{e}(R)>{\cal J}_{E}(z_{0})$ for $%
R>1000a_{0}$ between two identical atoms.

We will now discuss how this interaction Hamiltonian can be used for
logic gates. We first point out that this interaction resembles the
quantum gate implementation using the Heisenberg spin (exchange)
interaction \cite{loss} $H_{J}=J(t)\,\vec{\sigma}_{1}\cdot \vec{\sigma}_{2}$, which is
known to be universal. For $\int_{0}^{T}dt(J/\hbar )=\pi /4({\rm mod}2\pi )$%
, its unitary evolution operator creates a swap gate
\[
U_{{\rm swap}}(T)|i\rangle _{1}|j\rangle _{2}=\exp (-i{\frac{\pi }{4}}%
)|j\rangle _{1}|i\rangle _{2}.
\]
which in turn can be use to generate XOR (controlled-NOT) gates by
incorporating single bit operations \cite{loss}. However, our interaction
Hamiltonian includes an anisotropic term. Fortunately, we can borrow a
de-coupling technique developed in NMR \cite{abragham} to effect the
conversion of $\vec{\sigma}_{q}\cdot \vec{\sigma}_{h}\rightarrow \sigma
_{qz}\cdot \sigma _{hz}$, which is also universal. In fact, the phase gate
\cite{loss} in terms of these operators is simply
\[
U_{{\rm phase}}=e^{i(\pi /4)\sigma _{1z}\sigma _{2z}}\times e^{i(\pi
/4)\sigma _{1z}}\times e^{i(\pi /4)\sigma _{2z}},
\]
from which $U_{{\rm XOR}}$ can be easily made \cite{kimble,barenco}.
Furthermore, the swap gate, which we will require, can be made according to
\[
U_{{\rm swap}}(1\!\!\leftrightarrow \!\!2)=U_{{\rm XOR}}(1,2)U_{{\rm XOR}%
}(2,1)U_{{\rm XOR}}(1,2),
\]
where $U_{{\rm XOR}}(i,j)$ denotes a C-NOT with $i$ as the control bit
operating on $%
j$. The necessary de-coupling is achieved through a ``stirring'' radio
frequency field acting only on the $h$-atom\cite{abragham}, and is most
easily discussed in the context of the following model Hamiltonian
\begin{eqnarray}
H_{S}(t) &&=\hbar \omega _{1}\sigma _{1z}+\hbar \omega _{2}\sigma
_{2z}+\Omega _{S}(\sigma _{2+}e^{-i\omega _{S}t}+h.c.)  \nonumber \\
&&+\gamma _{e}(R)[\vec{\sigma}_{1}\cdot \vec{\sigma}_{2}-3\sigma _{1z}\sigma
_{2z}(\hat{R}\cdot \hat{z})^{2}],
\end{eqnarray}
where $\omega _{S}$ is the frequency of the stirring field,
and $\Omega _{S}$ is the Rabi frequency
of the stirring field. By analyzing this system in
the rotating frame defined by $U_{{\cal R}}=e^{i\omega _{S}t\sigma _{2z}}$,
we obtain \cite{abragham} $U_{{\cal R}}^{+}\sigma _{2\pm }U_{{\cal R}%
}\rightarrow \sigma _{2+}e^{\pm i\omega _{L}t},$ and invoking a rotating
wave approximation, the desired result is obtained,
\begin{eqnarray}
H_{S}^{{\rm eff}} &&\approx \gamma _{e}(R)[1-3(\hat{R}\cdot \hat{z}%
)^{2}]\sigma _{1z}\sigma _{2z}  \nonumber \\
&&+\hbar \,\omega _{1}\sigma _{1z}+\hbar (\omega _{2}-\omega _{S})\sigma
_{2z}+\Omega _{S}(\sigma _{2+}+\sigma _{2-}).
\end{eqnarray}
Although there are unwanted single atom terms in the second
line of this Hamiltonian, they can be easily
compensated with one-bit rotations. For our system,
a similar procedure yields the following effective
Hamiltonian,
\begin{eqnarray}
{\cal H}_{{\rm eff}}(\vec{R}) &\approx &[{\cal J}_{E}(z_{0})+\gamma
_{e}(R)-3\gamma _{e}(R)(\hat{R}\cdot \hat{z})^{2}]\sigma _{qz}\sigma _{hz}
\nonumber \\
&=&J_{E}(z_{0})\,\sigma _{qz}\sigma_{hz}.  \label{e4}
\end{eqnarray}
Interestingly, we note that the spin and spatial dependence
of the operators factorizes. This implies that coherent spin-spin
interactions only require that the motional states of the atoms remain
unchanged---the atoms are not necessarily required to be in the
ground state $|0\rangle$ of their respective trapping potential.
This particular feature of our proposal will be
discussed in detail elsewhere. At $%
z_{0}>R_{0}$, the effective spin-spin interaction strength is
\[
J_{E}(z_{0})\approx \langle \gamma _{e}(R)[1-3(\hat{z}\cdot \hat{R}%
)^{2}]\rangle ,
\]
which for atoms in the ground states $|0\rangle$ previously described
is readily evaluated
\begin{eqnarray}
&&\langle {\frac{1}{R^{3}}}[1-3(\hat{z}\cdot \hat{R})^{2}]\rangle   \nonumber
\\
&=&{\frac{1}{\sqrt{2\pi a_{z}^{2}}}}{\frac{1}{2a_{r}^{4}}}\int_{-\infty
}^{\infty }dz\exp \left( -{\frac{(z-z_{0})^{2}}{2a_{z}^{2}}}\right)
\nonumber \\
&&\left[ 2|z|-(a_{r}^{2}+z^{2}){\frac{\sqrt{2\pi }}{a_{\rho }}}\exp \left( {%
\frac{z^{2}}{2a_{r}^{2}}}\right) {\rm erfc}\left( {\frac{|z|}{\sqrt{2}a_{r}}}%
\right) \right] ,  \nonumber
\end{eqnarray}
where
erfc(.) is
the complementary error function. The geometry
of the system of two interacting spins are illustrated in
Fig. \ref{fig2}. The result of the effective interaction
is shown in Figure \ref{fig3},
we note that $J_{E}(z_{0})$
is in the kHz range for a distance of $1000a_{0}$ ($\sim50$ nm),
which will be more than adequate for gate operations for
atoms trapped in far off-resonant optical lattices.

The principle challenge in implementing this scheme is in providing the
appropriate confining potentials for the atoms. On one hand, the trapping
potentials for the individual atoms need always be distinguishable in order
to maintain identifiable qubits. On the other hand, as we can see from
Figure 2, the atoms need to be in close proximity ($\sim50$ nm) in
order for an appreciable interaction rate even for this `long-range'
potential. Previous proposals also requiring small inter-atomic spacings
have suggested spin-dependent traps created by optical lattices with
polarization gradients \cite{km,caves,zoller}.
Because of the nature of these types of traps, the
types of atomic manipulations are rather restricted, and hence scalability is
difficult.

To circumvent this complication, we will use two different atomic species,
one for the (stationary) quantum register, and one for the quantum header
atom. Each species of atom will be separately trapped by different laser
fields. By appropriate choice of atom and frequency of the trapping fields,
we can make these traps essentially independent. For a concrete example,
consider a quantum register consisting of an array of single atoms (type $q$
for qubit) trapped in 3D standing wave formed by interfering laser field of
a CO$_{2}$ lasers (wavelength $\lambda _{\rm CO_{2}}\approx 10.6$ $\mu $m)
\cite{co2}. The qubits will be separated by $\lambda _{\rm CO_{2}}/2$ which is
more than enough to allow individual addressing, and at this separation, the
long range Casimir-Polder interaction is negligible \cite{mircea}. The
trapping details are discussed in the appendix, but we point out that the
potential $V$ is very well approximated by the dc-polarizability of the atom
$\alpha \left( 0\right) $and the laser electric field amplitude $E$ as $%
V=-\alpha (0)E^{2}/2$. The relevant parameters are tabulated for
alkali atoms in Table \ref{table1}.

A separate laser field provides confinement for the header atom (of a
different type atom, $h$). By
choosing a trapping wavelength somewhat closer to the atomic resonance of $h$
(and detuned to the blue of the resonance), we can provide a potential which acts
principally on the $h$ atom. The potential depth for this trap is given by $V_{%
{\rm max}}=\hbar{\Omega _{L}^{2}/4\delta _{L}}$, with $\Omega _{L}$ the Rabi
frequency of the laser, and $\delta _{L}=\omega _{L}-\omega _{0}$ the
detuning.  The $h$ atom is also affected by the far off-resonant CO$_{2}$
laser field of course, but we can arrange for the off-resonant blue
detuned field to dominate the far off-resonant CO$_{2}$ laser potential
by suitable choice of atoms.
Trapping parameters for this case are listed in Table \ref
{table2}. The quantum register atoms (type $q)$ will also be affected, at
some level, by the blue lattice, but the detuning between the blue field and
the $q$ atoms will be much larger, so the potential will be
dominated by the CO$_{2}$ laser field for the $q$ atoms.

Gate operations in this system can be achieved in a three step process
requiring quantum state swapping between the quantum bits and the header
atom. To execute a gate operation between two qubits $q_{i}$ and $q_{j}$, we
first translate the header atom $h$ to the location of $q_{i}$ and perform
a state swap $q_{i}\!\!\leftrightarrow \!\!h.$  The header atom is then
translated to site $q_{j}$ and the gate operation between $h(q_{i})$ and $%
q_{j}$ is performed. Finally the header atom is translated  back to $q_{i}$
to and the state swap is repeated. The header atom effectively acts as an
quantum bus between the qubits, and in this sense our scheme shares certain
features with the quantum gear machine proposed by DiVincenzo \cite{gear}.

Single-bit operations can be realized either by directly addressing the
individual qubits $q_{i}$, or, alternatively, we can use the header atom as
a mediator. The latter option may be easier in some cases than
the direct spatial selection of
$q_{i}$ because the $h$ atoms can be sparsely distributed and have different
resonance level structures. The single bit operation will again be a three
step process: 1) perform a  state swap  between  $q_{i}$ and $h$,  2) perform
the arbitrary qubit operation on  $h$,  3) repeat the state swap between  $h$
and $q_{i}$.

In considering the ultimate scalability of this, and other lattice-based schemes,
it is necessary to compare the characteristic intrinsic decoherence time of
the system to the gate time {\em plus} the transport time of the moving
atoms \cite{zol}.  Additionally, the transport of the moving atoms
(the h-type atom in this case) must be adiabatic such the motional state
of the atom remains unchanged.  This latter condition implies constraints
on the magnitude of the motion, which we can estimate using perturbation
theory.  Consider the Hamiltonian for a one dimenstional harmonically
trapped particle with
mass $M$ and trap frequency $\omega_t$ subjected to a force $F(t)$,
\begin{eqnarray}
H={p^2\over 2M}+{1\over 2}M\omega_t^2q^2-q F(t),
\end{eqnarray}
adiabaticity condition for the header qubit translation requires
its motional state wave-function to be essentially unchanged.
This problem is equivalent to the problem of a translating
SHO potential $M\omega_t^2[q-q_0(t)]^2/2$
[with $F(t)=M\omega_t^2q_0(t)$] up to a deterministic phase factor
due to $M\omega_t^2q_0^2(t)/2$.
Calculating the probability for excitation out of the ground state
is a standard textbook problem and the result to first order is
\begin{eqnarray}
p_1^{(1)}={{1\over 2}M(\delta v)^2 \over \hbar\omega_t}\times
\exp(-\omega_t\tau),
\end{eqnarray}
for a time dependent force $F(t)=(F_0\tau/\omega)/(\tau^2+t^2)$.
${1\over 2}M(\delta v)^2$ is the energy gained from the impulse
 $M\delta v=\int_{-\infty}^{\infty}F(t)dt$ of the force.
We see that adiabaticity is maintained even after
the translating atom gains a very large speed, but satisfying
the condition $\omega_t\tau>\!>1$, i.e. a force to be slowly
turning on and off compared with the harmonic trap period.
Similar conclusions are reached for an initial coherent
motional state wave-packet. This condition effectively then
puts no constraint on the header atom speed, contrary
to the strong conditions as obtained in Ref. \cite{zol}.
For our problem, creative pulse shape design will allow
the header atom to be adiabatically transported over
many qubits within the single photon scattering
coherence time.

Our discussion thus far has been limited to alkali atoms with no nuclear
spin (e.g. $^{78}$Rb). When the nuclear spin $I$ is nonzero,
the atomic spin takes on values $F=I\pm {1/2}$ and we must include
the hyperfine interaction $V_{%
{\rm hf}}\sim \,a_{{\rm hf}}\vec{\sigma}\cdot \vec{\sigma}^{n}$.
The spin-spin interaction becomes considerably richer in detail.
However, if a strong Zeeman interaction is applied using a uniform
magnetic field, the resulting two manifolds of Zeeman states
correspond roughly to the electronic spin up/down
such that the good basis becomes $%
|I,S,I_{z},S_{z}\rangle $ \cite{walker}. Alternatively, we could chose an
atom with no nuclear spin such as $^{78}$Rb (radioactive lifetime about 20
minutes).

In summary we have proposed a new quantum computing implementation
with trapped atomic
spins. Utilizing dual optical lattices for two different type of atoms
provides a novel method to control the binary interaction between any pair
of qubits. In addition, our proposal, being based on the periodic structure
of optical lattices, is readily scalable, and in particular, redundant
parallel processing of information can be implemented using multiple $h$%
-type atoms operating on repetitive blocks of $q$-type atoms. This may be
useful in implementing error correction \cite{error}, concatenated coding,
and fault tolerant computing \cite{fault}.

We thank Drs. J. Cirac and P. Zoller for enlightening discussions. L. Y.
also thanks Dr. T. Walker and Dr. DiVincenzo for helpful communications.
We thank Dr. Z. T. Lu for information about nuclear spin 0 alkali isotopes.
This work is supported by the ARO/NSA grant DAA55-98-1-0370 and by the ONR research
grant No. 14-97-1-0633.

\appendix
\section{Possible trap parameters}

\begin{table}[tbp]
\caption{Parameters for different alkali-metal atoms inside a CO$_2$ lattice
with intensity $I=10^6$ (watts/cm$^2$).
For the `red' CO$_2$ lattice, the maximum level shift is
$V_{{\rm max}}= \alpha(0) E_0^2/4$.
At an intensity of $\sim 10^6$ (watts/cm$^2$), the single photon scattering
rate can provide decoherence times of many minutes. Assuming
a harmonic approximation, the oscillation frequency $\nu_{{\rm osc}}$ inside
the CO$_2$ trap is
$\nu_{{\rm osc}}=2\sqrt{V_{{\rm max}} E_R^{\rm CO_2}}$ with $%
E_R^{\rm CO_2}=h/(2M\lambda_{\rm CO_2}^2)$ the recoil energy (in Hz) for
emitting/absorbing a CO$_2$-photon. The Lamb-Dicke parameter $\eta_0=k_0 a_{%
{\rm osc}}=\sqrt{E_R^0/\nu_{{\rm osc}}}$ ($\eta_{\rm CO_2}$) measures the trap
ground state size $a_{{\rm osc}}$ in terms of the
resonant wavelength $\lambda_0$ (CO$_2$ laser $\lambda_{\rm CO_2}$). }
\label{table1}
\begin{tabular}{ccccccc}
& Li & Na & K & Rb & Cs &  \\
\tableline M & 6.9 & 23 & 39 & 87 & 133 &  \\
$\alpha(0)$ ($a_0^3$) & 159.2 & 162 & 292.8 & 319.2 & 402.2 &  \\
$V_{{\rm max}}$ (MHz) & 181 & 185 & 334 & 364 & 458 &  \\
$\nu_{{\rm osc}}$ (kHz) & 432 & 239 & 247 & 172 & 156 &  \\
$a_{{\rm osc}}$ ($a_0$) & 778 & 573 & 433 & 347 & 295 &  \\
$\lambda_0$ (nm) & 670 & 589 & 766 & 780 & 852 &  \\
$E_R$ (kHz) & 64 & 25 & 8.7 & 3.7 & 2 &  \\
$\eta_0$ & 0.39 & 0.32 & 0.19 & 0.15 & 0.11 &  \\
$\eta_{\rm CO_2}$ & 0.025 & 0.018 & 0.014 & 0.011 & 0.009 &
\end{tabular}
\end{table}

\begin{table}[tbp]
\caption{Parameters for a `blue' lattice with
$\Omega_L\sim 1.6 \times 10^{10}$ (Hz) ($\sim$ laser power of 10 kw/cm$^2$),
$\protect\gamma=10^7$ (Hz), and $%
\protect\delta_L=2\times 10^{12}$ (Hz).
For the near resonant `blue' lattice on the $h-$type atoms,
the effective single
photon scattering rate is approximately $\gamma_{{\rm eff}}=\eta^2 {\frac{%
\Omega_L^2}{4\delta_L^2}}\gamma$. We see as indicated in Table \ref{table2}
the confining frequency $\nu_{{\rm osc}}$ is indeed much larger than that of
CO$_2$ laser (on $q$-type atoms).}
\label{table2}
\begin{tabular}{ccccccc}
& Li & Na & K & Rb & Cs &  \\
\tableline M & 6.9 & 23 & 39 & 87 & 133 &  \\
$\nu_{{\rm osc}}$ (kHz) & 4061 & 2530 & 1494 & 982 & 727 &  \\
$a_{{\rm osc}}$ ($a_0$) & 254 & 176 & 176 & 145 & 137 &  \\
$\eta$ & 0.13 & 0.1 & 0.076 & 0.06 & 0.05 &  \\
$\gamma_{{\rm eff}}$ (Hz) & 2.5 & 1.6 & 0.9 & 0.6 & 0.5 &
\end{tabular}
\end{table}

\begin{figure}[tbp]
\centerline{\epsfig{file=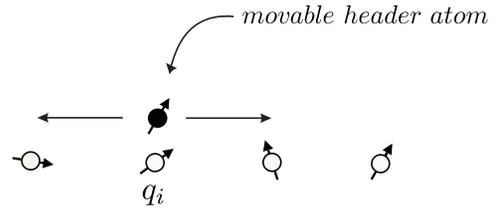,width=2.5in}\\[12pt]}
\caption{A one dimensional illustration. The $q-$type atom array
are trapped in a red periodic CO$_{2}$ laser lattice.
The movable header atom is trapped in a blue lattice.}
\label{fig1}
\end{figure}

\begin{figure}[tbp]
\centerline{\epsfig{file=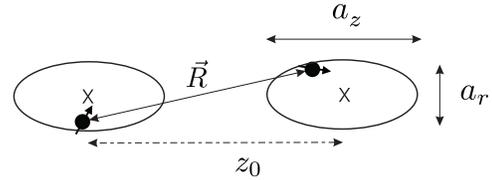,width=2.5in}\\[12pt]}
\caption{The geometry of interacting header atom and
qubit atom pair. The large ellipses denote trap ground
states with trap centers crossed and separated by $z_0$.
Solid circles with arrow heads denote
electron spins separated by $\vec R$.
}
\label{fig2}
\end{figure}

\begin{figure}[tbp]
\centerline{\epsfig{file=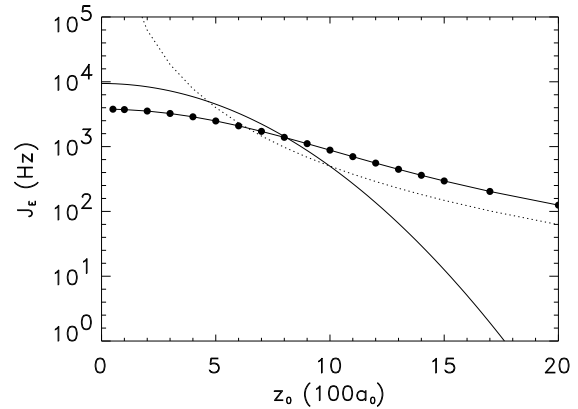,width=3.in}\\[12pt]} \caption{The
solid line denotes the exchange interaction
${\cal J}_{E}$ assuming a absolute difference
of the $|a_{S}-a_T|=100$ ($a_0$), while the dots are are numerical
results of the averaged spin dipole interaction strength $J_{E}$
for $a_{qr}=a_{qz}=400a_{0}$ and $a_{hr}=a_{hz}=100a_{0}$. The
dotted line represents the simple $1/z_{0}^{3}$ dependence. As
pointed out in the text, the exchange interaction will be
significantly suppressed for the case where the two atoms
are two different species. } \label{fig3}
\end{figure}

\end{document}